\documentclass[aps,nofootinbib,noshowpacs,showkeys,preprintnumbers,amsmath,amssymb]{revtex4}
\usepackage{graphicx,color}
\usepackage{graphics}
\usepackage{rotating}
\usepackage{amssymb}

\usepackage{color}

\usepackage{epsfig}
\usepackage{dcolumn}
\usepackage{bm}
\usepackage{multirow}

\def\be{\begin{equation}}
\def\ee{\end{equation}}
\def\ba{\begin{eqnarray}}
\def\ea{\end{eqnarray}}
\def\bea{\begin{eqnarray}}
\def\eea{\end{eqnarray}}
\def\bes{\begin{subequations}}
\def\ees{\end{subequations}}
\def\bear{\begin{array}}
\def\eear{\end{array}}

\newcommand{\A}{{\mathcal{A}}}

\newcommand{\ta}{{\widetilde a}}

\newcommand{\tK}{{\widetilde K}}
\newcommand{\td}{{\widetilde d}}

\newcommand{\MSbar}{\overline{\rm MS}}

\begin{document}

\title{Evaluation of Bjorken polarised sum rule with a renormalon-motivated approach}
\author{C\'esar Ayala$^a$}
\email{c.ayala86@gmail.com}
\author{Camilo Castro-Arriaza$^b$}
\email{camilo.castroa@sansano.usm.cl}
\author{Gorazd Cveti\v{c}$^b$}
\email{gorazd.cvetic@gmail.com}
\affiliation{$^a$Instituto de Alta Investigaci\'on, Sede La Tirana, Universidad de Tarapac\'a, Av.~La Tirana 4802, Iquique, Chile}
\affiliation{$^b$Department of Physics, Universidad T{\'e}cnica Federico Santa Mar{\'\i}a, Avenida España 1680, Valpara{\'\i}so, Chile}

\date{\today}

\begin{abstract}
We use the known renormalon structure of Bjorken polarised sum rule (BSR) ${\overline \Gamma}_1^{p-n}(Q^2)$ to evaluate the leading-twist part of that quantity. In addition, we include $D=2$ and $D=4$ Operator Product Expansion (OPE) terms and fit this expression to available experimental data for inelastic BSR. Since we use perturbative QCD (pQCD) coupling, which fails at low squared spacelike momenta $Q^2 \lesssim 1 \ {\rm GeV}^2$ due to Landau singularities, the fit is performed for $Q^2 \geq Q^2_{\rm min}$ where $Q^2_{\rm min} \approx (1.7 \pm 0.3) \ {\rm GeV}^2$.  Due to large BSR experimental uncertainties, the extracted value of the pQCD coupling has very large uncertainties, especially when $Q^2_{\rm min}$ is varied. However, when we fix the pQCD coupling to the known world average values, the $D=2$ and $D=4$ residue parameters can be determined within large but reasonable uncertainties.
\end{abstract}
\maketitle

\section{Introduction}
\label{sec:intr}

The polarised Bjorken sum rule (BSR)  ${\Gamma}_1^{p-n}(Q^2)$ \cite{BjorkenSR} is the difference of the first moment of the spin-dependent structure functions of proton and neutron. Due to its spacelike and isovector nature it has a relatively simple form of OPE. Experiments with scattering of polarised leptons on polarised targets give us measured values of the inelastic BSR  ${\overline \Gamma}_1^{p-n}(Q^2)$. They have been performed over a large interval of $Q^2$, $0.02 \ {\rm GeV}^2 < Q^2 < 5 \ {\rm GeV}^2$, at  CERN \cite{CERN}, DESY \cite{DESY}, SLAC \cite{SLAC}, and at various experiments at the Jefferson Lab \cite{JeffL1,JeffL2,JeffL3,JeffL4,JeffL5}. These data points still have significant statistical and systematic uncertainties.

The theoretical evaluation of the inelastic BSR ${\overline \Gamma}_1^{p-n}(Q^2)$ is performed usually by using truncated OPE, and the leading-twist (i.e., the dimension $D=0$ part) is evaluated usually by using truncated perturbation series or specific variants thereof \cite{JeffL1,JeffL2,JeffL4,ACKS,BSRPMC}. In the $D=0$ part of BSR, the coefficients at powers of the perturbative QCD (pQCD) coupling, $a(\mu^2) \equiv \alpha_s(\mu^2)/\pi$, are explicitly known up to power $a^4$. Then the resulting truncated OPE (usually truncated at the dimension $D=2$ or $D=4$ term) is fitted to the experimental data, and the (effective) $D=2$, and possibly $D=4$, coefficients are determined, as well as the resulting quality of this fit.

Since the experimental data are available also at low momenta $Q^2 \lesssim 1 \ {\rm GeV}^2$ where the pQCD couplings $a(Q^2)$ in general have Landau singularities, which do not reflect the holomorphic properties of spacelike quantities such as ${\overline \Gamma}_1^{p-n}(Q^2)$, approaches with holomorphic (analytic) QCD couplings [$a(Q^2) \mapsto \mathcal{A}(Q^2)$] have been used in such regimes \cite{ACKS,KZ}, as well as specific low-$Q^2$ models \cite{JeffL2,LFH,LFHBSR}.

In this work we return to pQCD to evaluate the leading-twist part $d(Q^2)$ of BSR, by applying a specific renormalon-motivated approach \cite{renmod}, where we use the relatively good knowledge of the renormalon structure of $d(Q^2)$ and the knowledge of several first coefficients of the perturbation series of $d(Q^2)$. Using this renormalon-motivated resummation of the $D=0$ part in the OPE, we then perform fits to the experimental data for the inelastic BSR, extract the effective $D=2$ and $D=4$ coefficients of the OPE, and comment on the quality of the fit and on the regime of validity of this resummed pQCD approach.  

In Sec.~\ref{sec:BSR} we write down the theoretical (OPE) expressions of BSR. In Sec.~\ref{sec:RS} we present \textcolor{black}{a specific form of description of the renormalon structure} for the canonical $D=0$ part, $d(Q^2)$, of BSR. \textcolor{black}{In Sec.~\ref{sec:res} the corresponding resummation expression of $d(Q^2)$ is derived.} In Sec.~\ref{sec:RSch} we explain the fixing of the renormalisation scheme. And finally in Sec.~\ref{sec:fit} we present the fitting procedure, show the results for the extracted parameters, and make conclusions.

\section{Theoretical expressions of Bjorken sum rule}
\label{sec:BSR}

The polarised Bjorken sum rule (BSR), ${\overline \Gamma}_1^{p-n}$, is the difference between the polarised structure functions $g_1$ integrated over the $x$-Bjorken interval
\be
{\overline \Gamma}_1^{p-n}(Q^2)=\int_0^{1-0} dx \left[g_1^p(x,Q^2)-g_1^n(x,Q^2) \right]\ .
\label{BSRdef}
\ee 
The bar over $\Gamma_1$ denotes that we take here only the inelastic part (i.e., $x<1$). The inelastic BSR ${\overline \Gamma}_1^{p-n}(Q^2)$ has been extracted at various values of $Q^2 \equiv -q^2 > 0$, from various experiments \cite{CERN,DESY,SLAC,JeffL1,JeffL2,JeffL3,JeffL4,JeffL5}.
The theoretical Operator Product Expansion (OPE) for this quantity has the form
\cite{BjorkenSR,BSR70}
\be
\label{BSROPE}
{\overline \Gamma}_1^{p-n, \rm {OPE}}(Q^2)= {\Big |}\frac{g_A}{g_V} {\Big |} \frac{1}{6}(1 - d(Q^2) - \delta d(Q^2)_{m_c})+\sum_{i=2}^\infty \frac{\mu_{2i}(Q^2)}{Q^{2i-2}}\ ,
\ee
where we take $|g_A/g_V|=1.2754$ (consistent with PDG 2020 \cite{PDG2020}) for the ratio of the nucleon axial charge, $d(Q^2) = a(Q^2) + {\cal O}(a^2)$ is the canonical massless pQCD part \textcolor{black}{(with $N_f=3$), $\delta d(Q^2)_{m_c}$ are the corrections to the decoupling due to $m_c \not= \infty$,} and $\mu_{2i}/Q^{2i-2}$ are $D \equiv (2 i -2) \geq 2$ contributions. The term proportional to  $|g_A/g_V|$ is the total leading-twist (LT) contribution ($D=0$).

The canonical part, $d(Q^2)$, has perturbation expansion in powers of  $a \equiv a(Q^2) \equiv {\alpha_s}(Q^2)/\pi$
\bea
d(Q^2)_{\rm pt} & = & a + d_1 a^2 + d_2 a^3 + d_3 a^4 + {\cal O}(a^5).
\label{dptMSb}
\eea
In the $\MSbar$ scheme we have $d_j = d_j^{\MSbar}$ and $a= a^{\MSbar}(Q^2)$. Hereafter, we will consider $N_f=3$, for $Q^2 < (2 m_c)^2$ regime. The coefficients $d_j^{\MSbar}$ ($j=1,2,3$) were obtained in  \cite{GorLar1986,LarVer1991,BaiCheKu2010}. In any other scheme, the series is then also known up to $a^4$ (e.g., cf.~App.~A of \cite{ACKS}).

The $D=2$ OPE coefficient $\mu_4$ has known $Q^2$-dependence
\be
\frac{\mu_4(Q^2)}{Q^2} = \frac{M_N^2}{9} \frac{\left[ A+ 4 {\bar f}_2 \; {a}(Q^2)^{{k_1}} \right]}{Q^2},
\label{BSRD2} \ee
where ${k_1}=32/81$ is the anomalous dimension \cite{Kawetal1996}, $M_N =0.9389$ GeV is the  nucleon mass, and the constant $A = (a_2^{p-n} + 4 d_2^{p-n}) \approx 0.063$ contains the (twist-2) target correction $a_2^{p-n} \approx 0.031$ and a twist-3 matrix element $d_2^{p-n} = \int dx x^2 (2 g_1^{p-n} + 3 g_2^{p-n})$ $\approx 0.008$. The parameter ${\bar f}_2$ will be determined by the fit, as will be the $D=4$ OPE coefficient $\mu_6$ which will be considered $Q^2$-independent.

The next coefficient $d_4^{\MSbar}$ in the perturbation series can be estimated, for example, by the ECH method \cite{KS}
\be
d_4^{\MSbar} = d_4^{\MSbar}({\rm ECH}) \pm 32.8 \approx 1557.4 \pm 32.8.
\label{d4est}
\ee
The uncertainty $\pm 32.8$ was estimated here by the following reasoning:
when $d_4^{\MSbar}=1557.4-32.8=1524.6$, we obtain in the preferred renormalisation scheme [$c_2=9.$ and $c_3=20.$, cf.~Eq.~(\ref{RGE2})] for ${\cal B}[\td](u)$ the disappearance of the u=-2 UV renormalon, i.e., ${\tilde d}^{\rm UV}_2=0$., cf.~Sections \ref{sec:RS} and \ref{sec:RSch}.

The part $\delta d(Q^2)_{m_c}$ in Eq.~(\ref{BSROPE}) appears due to the fact that the charm quark in our approach does not fully decouple, i.e., $m_c \not= \infty$ (in our fit we will have: $1.71 \ {\rm GeV}^2 < Q^2 < 4.74 \ {\rm GeV}^2$). This part is known up to ${\cal O}(a^2)$, it was obtained in \cite{Blumetal} and can be written as
\be
\delta d(Q^2)_{m_c} =
\frac{1}{6} \left[ \ln \left( \frac{Q^2}{m_c^2} \right) -
  2 C^{\rm mass.,(2)}_{\rm pBJ}\left( \frac{Q^2}{m_c^2} \right) \right] a(Q^2)^2 +
{\cal O}(a^3),
\label{deld} \ee
where $m_c \approx 1.67$ GeV is the pole mass of the charm quark \cite{PDG2023}, and the function $ C^{\rm mass.,(2)}_{\rm pBJ}(\xi)$ was obtained in \cite{Blumetal} (cf.~also App.~E of \cite{ACKS}). We note that the logarithmic term above is obtained from the difference $a(Q^2;N_f=4)-a(Q^2;N_f=3)$$=(1/6) \ln(Q^2/m_c^2) a^2 + {\cal O}(a^3)$.

\section{Renormalon structure of $d(Q^2)$}
\label{sec:RS} 

According to the approach of Ref.~\cite{renmod}, to account for the renormalon structure and obtain the characteristic function of $d(Q^2)$, it is important to construct first \textcolor{black}{a modified, reorganised, expansion of $d(Q^2)$}. The above power series (\ref{dptMSb}), \textcolor{black}{can be written at any chosen renormalisation scale $\mu^2$ (and in any renormalisation scheme),} and can be subsequently reorganised in logarithmic derivatives ${\ta}_{n+1}$
\begingroup \color{black}
\be
{\ta}_{n+1}(\mu^2) \equiv \frac{(-1)^n}{n! \beta_0^n} \left( \frac{d}{d \ln \mu^2} \right)^n {a}(\mu^2) \qquad (n=0,1,2,\ldots),
\label{tan} \ee
[note: ${\ta}_{n+1}(\mu^2) = {a}(\mu^2)^{n+1} + {\cal O}({a}^{n+2})$]\footnote{\textcolor{black}{Here, $\beta_0=(11 - 2 N_f/3)/4 = 9/4$ ($N_f=3$) is the one-loop beta-coefficient, i.e., $d a(\mu^2)/d \ln \mu^2 = - \beta_0 a(\mu^2)^2 (1 + {\cal O}(a))$.}}
and we obtain
\bes
\label{dptdlpt}
\bea
d(Q^2) &=& a(\kappa Q^2) + d_1(\kappa) a(\kappa Q^2)^2 + \ldots + d_n(\kappa) a(\kappa Q^2)^{n+1} + \ldots
\label{dpt} \\
& = & a( \kappa Q^2) + {\td}_1(\kappa) \; {{\ta}}_2(\kappa Q^2) + \ldots + {\td}_n(\kappa) \; {{\ta}}_{n+1}(\kappa Q^2) + \ldots,
\label{dlpt} \eea \ees
where $\kappa \equiv \mu^2/Q^2$ ($0 < \kappa \lesssim 1$) and $\mu^2$ is a chosen renormalisation scale. Since $d(Q^2)$ is an observable, it is independent of $\kappa$.
\endgroup
By the use of the renormalisation group equation (RGE), in a chosen renormalisation scheme, we can relate the new coefficients $\td_n(\kappa)$ with the original ones $d_n(\kappa), d_{n-1}(\kappa), \ldots$. \textcolor{black}{Also the corresponding inverse relations can be constructed, i.e., $d_n(\kappa)$ as a combination of $\td_n(\kappa), \td_{n-1}(\kappa), \ldots$. Stated otherwise, the sequence of coefficients $\td_0(\kappa)(=1)$, $\td_1(\kappa), \ldots, \td_n(\kappa)$ contains all the information that the sequence of coefficients  $d_0(\kappa)(=1)$, $d_1(\kappa), \ldots, d_n(\kappa)$ contains; and viceversa.}

\begingroup \color{black}
In complete analogy with the Borel transform ${\cal B}[d](u;\kappa)$, which generates the coefficients $d_n(\kappa)$,
\bes
\label{calB}
\bea
{\cal B}[d](u;\kappa)_{\rm ser} &\equiv& 1 + \frac{{d}_1(\kappa)}{1! \beta_0} u + \ldots + \frac{{d}_n(\kappa)}{n! \beta_0^n} u^n + \ldots \; \Leftrightarrow
\label{Bdser} \\
d(Q^2) & = & \frac{1}{\beta_0} \int_0^{\infty} du \exp \left[- \frac{u}{\beta_0 a(\kappa Q^2)} \right] {\cal B}[d](u; \kappa),
\label{dBorsum} \eea \ees  
we define the corresponding Borel transform ${\cal B}[\td](u;\kappa)$ that generates the new coefficients $\td_n(\kappa)$, i.e., in the power series of ${\cal B}[d](u;\kappa)$ we replace $d_n(\kappa) \mapsto {\td}_n(\kappa)$
\be
{\cal B}[\td](u;\kappa)_{\rm ser} \equiv 1 + \frac{{\td}_1(\kappa)}{1! \beta_0} u + \ldots + \frac{{\td}_n(\kappa)}{n! \beta_0^n} u^n + \ldots.
\label{Btdser} \ee
\endgroup
It turns out that this \textcolor{black}{new Borel transform} has a very simple, one-loop type, form of dependence on the renormalisation scale $\mu^2$ 
\bes
\label{kapdep}
\bea
\frac{d}{d \ln \kappa} {\td}_n(\kappa) &=& n \beta_0 {\td}_{n-1}(\kappa) \quad \Rightarrow
\label{tdnkap} \\
\quad {\cal B}[\td](u; \kappa) &=& \kappa^u {\cal B}[\td](u).
\label{Btdkap} \eea \ees
\begingroup \color{black}
We point out that this $\kappa$-dependence follows from (exact) $\kappa$-independence of $d(Q^2)$, and that it is exact [in contrast to the case of the coefficients $d_n(\kappa)$ and the Borel transform ${\cal B}[d](u; \kappa)$, where it is valid only at the one-loop approximation].

In order to see how the relations (\ref{kapdep}) come about, we note that $d \ln \mu^2 = d \ln \kappa$, and that according to notations (\ref{tan}) we have the simple recursive relations
\be
\frac{d}{d \ln \kappa} {\ta}_n(\kappa Q^2) = (- \beta_0) n {\ta}_{n+1}(\kappa Q^2),
\label{rec} \ee
and thus the application of the derivative $d/d \ln \kappa$ on Eq.~(\ref{dlpt}) gives
\be
\frac{d}{d \ln \kappa} d(Q^2) = \sum_{n=1}^{\infty} \ta_{n+1}(\kappa Q^2) \left[(-\beta_0) n {\td}_{n-1}(\kappa)+ \frac{d}{d \ln \kappa} {\td}_n(\kappa) \right].
\label{dddln} \ee
Since $d(Q^2)$ is $\kappa$-independent, the relations (\ref{tdnkap}) follow directly from the relation (\ref{dddln}). Further, if we apply the derivative $d/d \ln \kappa$ on the quantity  ${\cal B}[\td](u; \kappa)$ in Eq.~(\ref{Btdser}), and take into account the obtained relations (\ref{tdnkap}), we obtain
\be
\frac{d}{d \ln \kappa}{\cal B}[\td](u; \kappa) = u {\cal B}[\td](u; \kappa),
\label{dBtddln} \ee
and then the relation (\ref{Btdkap}) follows immediately. This concludes the proof of the relations (\ref{kapdep}).
\endgroup

As a consequence, it can be shown \cite{renmod} that this Borel transform ${\cal B}[\td](u)$ has a structure very similar to the known \cite{BK1993,Renormalons} large-$\beta_0$ structure of the Borel ${\cal B}[d](u)$ \textcolor{black}{(cf.~also \cite{Kat1,Kat2})}
\bea
{\cal B}[\td](u) & = & \exp \left( {\tK} u \right) \pi {\Big \{}
{\td_{1}^{\rm IR}} \frac{1}{({1}-u)^{{\kappa_1}}} +
{\td_{2}^{\rm IR}} \frac{1}{({2}-u)} +
{\td_{1}^{\rm UV}} \frac{1}{({1}+u)} + 
{\td_{2}^{\rm UV}} \frac{1}{({2}+u)} {\Big \}},
\label{Btd5P}
\eea
Here, ${\kappa_1} = 1 - {k_1}$, where ${k_1}=32/81$ is the aforementioned anomalous dimension of the $D=2$ OPE term.\footnote{
The reason for this, as argued later in Eqs.~(\ref{ImBdpk}), lies in the fact that the corresponding renormalon ambiguity in the Borel-resummed quantity $d(Q^2)$ has the same $Q^2$-dependence $\sim a(Q^2)^{k_1}/Q^2$ as has the Bjorken $D=2$ OPE term Eq.~(\ref{BSRD2}).} 
The five parameters (${\tK}$ and the residues ${\td_{1}^{\rm IR}}, {\td_{2}^{\rm IR}}, {\td_{1}^{\rm UV}}, {\td_{2}^{\rm UV}}$) are determined by the knowledge of the first five coefficients $d_n$ (and thus $\td_n$), $n=0,1,2,3,4$. In Table \ref{tabrenmod1} we present the numerical values for these five parameters in the case of the (5-loop) $\MSbar$ scheme, and in the P44-scheme with $c_2=9.$ and $c_3=20.$ (that is explained later in Sec.~\ref{sec:RSch}).
\begin{table}
\caption{\footnotesize The values of ${\tK}$ and of the renormalon residues ${\td_j^{\rm X}}$ (X=IR,UV) for the five-parameter ansatz (\ref{Btd5P}) in the (5-loop) $\MSbar$ and in the 'P44' scheme with $c_2=9.$ \& $c_3=20.$, and with $c_2^{\MSbar}$($=4.47106$)  and $c_3^{\MSbar}$($=20.9902$) (for the schemes 'P44', see Sec.~\ref{sec:RSch}), when $d_4$ is taken such that it corresponds to the 5-loop $\MSbar$ value $d_4^{\MSbar}= 1557.43$ (as predicted by ECH). The last line is again for the case of 'P44' scheme with $c_2=9.$ \& $c_3=20.$, but  $d_4^{\MSbar}= 1557.43-32.84=1524.59$.} 
\label{tabrenmod1}
\begin{tabular}{r|rrrrr}
  scheme & ${\tK}$ & ${\td_{1}^{\rm IR}}$ & ${\td_{2}^{\rm IR}}$ &  ${\td_{1}^{\rm UV}}$ & ${\td_{2}^{\rm UV}}$ 
\\
\hline
\hline
$\MSbar$ (5-loop)  & -1.82336 & 7.81560 & -14.8199 & -0.0413348 & -0.0920349
\\
$\MSbar$ (P44)  & -1.83223 & 7.86652 & -14.9299 & -0.0444416 & -0.0776748
\\
$c_2=9.$ \& $c_3=20.$  (P44) & 0.450041 & 0.331813 & 0.231437 & -0.0809782 & -0.0964868
\\
\hline
$d_4^{\MSbar}=1524.6$ & 0.528239 & 0.276962 & 0.283465 & -0.100381 &  ${\cal O}(10^{-5})$
\\
\hline
\end{tabular}
\end{table}

\begingroup \color{black}
We point out that  ${\cal B}[\td][(u)$ in Eq.~(\ref{Btd5P}) is for the renormalisation scale choice $\kappa=1$ (i.e., $\mu^2 = Q^2$), and thus its expansion in powers of $u$ generates the coefficients $\td_n$ $[ \equiv \td_n(\kappa=1)$]. The Borel transform which generates $\td_n(\kappa)$ (for any chosen $\kappa$) is then, due to the relation (\ref{Btdkap}), equal to
\bea
{\cal B}[\td](u;\kappa) & = & \exp \left( ( \ln \kappa + {\tK}) u \right) \pi {\Big \{}
{\td_{1}^{\rm IR}} \frac{1}{({1}-u)^{{\kappa_1}}} +
{\td_{2}^{\rm IR}} \frac{1}{({2}-u)} +
{\td_{1}^{\rm UV}} \frac{1}{({1}+u)} + 
{\td_{2}^{\rm UV}} \frac{1}{({2}+u)} {\Big \}},
\label{Btd5Pkap}
\eea
We point out that, if we start, instead of the ansatz (\ref{Btd5P}) for ${\cal B}[\td](u)$, with the corresponding ansatz (\ref{Btd5Pkap}) for ${\cal B}[\td](u; \kappa)$ with a chosen $\kappa$, and determine the parameters in the aforedescribed way by using the information on the first five coefficients $\td_n(\kappa)$ ($n=0,1,2,3$), we obtain the very same values of the residues ${\td}_j^{\rm X}$ ($j=1,2$; X=IR, UV) and of the parameter $\tK$, which shows consistency of our approach. 
\endgroup

It can be shown \cite{renmod,ACT2023} that this ansatz for ${\cal B}[\td](u)$ implies the theoretically expected structure of the corresponding renormalon terms in the Borel ${\cal B}[d](u)$ of the canonical BSR $d(Q^2)$
\bes
\label{Btdpdp}
\bea
{\cal B}[\td](u) &=& \frac{A}{({p}\mp u)^{{\kappa}}} \;
\Rightarrow
\label{Btdp}
\\
{\cal B}[d](u) &=& \frac{B}{({p}\mp u)^{{\kappa}\pm {p} \beta_1/\beta_0^2}} \left[1 + {\cal O}({p}\mp u) \right],
\label{Bdp}
\eea \ees
where $\beta_0$ and $\beta_1$ are the one-loop and two-loop QCD $\beta$-coefficients (they are universal) appearing in the RGE
\bes
\label{RGE}
\bea
\frac{d {a}(Q^2)}{d \ln Q^2} &=& - \beta_0 {a}(Q^2)^2 - \beta_1 {a}(Q^2)^3 - \beta_2 {a}(Q^2)^4 - \ldots  
\label{RGE1} \\
& = &  - \beta_0 {a}(Q^2)^2 \left[ 1 + c_1 a + c_2 a^2 + \ldots \right].
\label{RGE2} \eea \ees
The renormalon ambiguity in the Borel-resummed expression of the IR renormalon term (\ref{Bdp}) has the following $Q^2$-dependence:
\bes
\label{ImBdpk}
\bea
\delta d (Q^2)_{p,\kappa} & \sim & 
\frac{1}{\beta_0} {\rm Im} \int_{+ i \varepsilon}^{+\infty + i \varepsilon} d u \exp \left( - \frac{u}{\beta_0 a(Q^2)} \right)
\frac{1}{(p - u)^{\kappa + p c_1/\beta_0}}
\label{ImBdpka} \\
& \sim & \frac{1}{(Q^2)^p} a(Q^2)^{1 -\kappa}  \left[ 1 + {\cal O}(a) \right].
\label{ImBdpkb}
\eea \ees
This means that the renormalon ambiguity corresponding to the IR-terms of the ansatz of the entire Borel $B[\td](u)$ Eq.~(\ref{Btd5P}) has the same $Q^2$-dependence as the $D=2$ and $D=4$ OPE terms of BSR Eqs.~(\ref{BSROPE}) and (\ref{BSRD2}) (we note that we take $\mu_6$ as $Q^2$-independent).

\begingroup \color{black}
\section{Resummation}
\label{sec:res}

If we know all the expansion coefficients ${\td}_n$ ($\leftrightarrow d_n$) of $d(Q^2)$, such as is the case in the ansatz Eq.~(\ref{Btd5P}), it turns out that we can resum the full expansion (\ref{dlpt}) of $d(Q^2)$ in the simple form \cite{renmod}
\be
d(Q^2)_{\rm res} = \int_0^\infty \frac{dt}{t} F_d(t) a(t Q^2),
\label{resFd} \ee
where $F_d$ is the characteristic function of $d(Q^2)$. We point out that: (a) $F_d(t)$ is constructed from the knowledge of the expansion coefficients $\td_n(\kappa)$ [or equivalently: $\leftrightarrow d_n(\kappa)$]; (b) $F_d(t)$ is independent of the renormalisation scale parameter $\kappa$ in a strong sense, i.e., this $\kappa$-independence does not involve any effective fixing of $\kappa$ to an optimal value. This means that the resummation (\ref{resFd}) is completely $\kappa$-independent, in the mentioned strong sense.

For clarity, we recapitulate here briefly the construction of $F_d(t)$. In the integrand of Eq.~(\ref{resFd}) we Taylor-expand $a(t Q^2)$ around $a(\kappa Q^2)=a(\mu^2) $ (where the relevant variable is the logarithm of the squared momenta)
\be
a(t Q^2) = a(\kappa Q^2) + (-\beta_0) \ln(t/\kappa) {\ta}_2(\kappa Q^2) + \ldots
+ (-\beta_0)^n \ln^n(t/\kappa) {\ta}_{n+1}(\kappa Q^2) + \ldots,
\label{Taylor} \ee
where we used the notation (\ref{tan}) for $\ta_{n+1}$. When we exchange the order of summation and integration in Eq.~(\ref{resFd}), and take into account that the perturbation expansion of $d(Q^2)$ in $\ta_{n+1}(\kappa Q^2)$ is that of Eq.~(\ref{dlpt}), we obtain the string of relations (requirements) for $F_d(t)$
\be
{\td}_n(\kappa) = (-\beta_0)^n \int_0^{\infty}
\frac{dt}{t} F_d(t) \ln^n \left( \frac{t}{\kappa} \right) \quad (n=0,1,2,\ldots).
\label{tdnSR} \ee
Now we multiply each of these relations by $u^n/(n! \beta_0^n)$ and sum over $n$, and on the right-hand side we exchange the order of summation and integration.\footnote{Explicitly, we use the summation formula $\sum_{n=0}^{\infty} (-1)^n w^n/n! = \exp(-w)$ for $w=u \ln(t/\kappa)$, i.e., $\exp(-w) = t^{-u} \kappa^u$.}
In this way, we arrive at the following relation:
\be
{\cal B}[\td](u; \kappa) =  \int_0^\infty \frac{dt}{t} F_d(t) t^{-u} \kappa^u.
\label{calB1} \ee
If we now take into account the relation (\ref{Btdkap}), the common factor $\kappa^u$ on both sides of Eq.~(\ref{calB1}) cancels out, thus all the $\kappa$-dependence cancels out, and we obtain
\be
{\cal B}[\td](u) =  \int_0^\infty \frac{dt}{t} F_d(t) t^{-u}.
\label{calB2} \ee
This means that ${\cal B}[\td](u)$ is Mellin transform of the (sought for) characteristic function $F_d(t)$; the latter is thus the inverse Mellin transform of ${\cal B}[\td](u)$
\be
F_d(t) = \frac{1}{2 \pi i} \int_{u_0 - i \infty}^{u_0+\infty} du \; {\cal B}[\td](u) t^u,
\label{Fd} \ee
where $u_0$ is any real value close to zero where the Mellin transform (\ref{calB2}) exists, i.e., in the BSR case we have $-1 < u_0 < +1$.

We wish to point out that this construction is based entirely on the (assumed) knowledge of the coefficients $\td_n(\kappa)$ of the expansion (\ref{dlpt}) at any chosen value of $\kappa$, and yet it gives us the characteristic function which is $\kappa$-independent, and thus the resummation Eq.~(\ref{resFd}) of $d(Q^2)$ is $\kappa$-independent.

In practical terms, the expression ${\cal B}[\td](u; \kappa)$, Eq.~(\ref{Btd5Pkap}), for any chosen renormalisation scale parameter value $\kappa$, is fixed on the knowledge of the first five expansion coefficients $\td_n(\kappa)$ ($n=0,1,2,3,4$), and gives us the values of the five parameters ${\td}_j^{\rm X}$ ($j=1,2$; X=IR, UV) and $\tK$ that are independent of the chosen value of $\kappa$, as pointed out in the previous Section. These same parameter values enter then also in ${\cal B}[\td](u)$ Eq.~(\ref{Btd5P}), where we have the relation ${\cal B}[\td](u; \kappa)= \kappa^u {\cal B}[\td](u)$ consistent with relation (\ref{Btdkap}). This approach then gives us the ($\kappa$-independent) characteristic function Eq.~(\ref{Fd}) and the resummation expression (\ref{resFd}) of $d(Q^2)$.

It is straightforward to see that the effect of the exponential factor $\exp(\tK u)$ in ${\cal B}[\td](u)$ of Eq.~(\ref{Btd5P}) reflects itself in a simple rescaling of the variable in the characteristic function $F_d(t)$ Eq.~(\ref{Fd})
\be
F(t)_d = G_d(t \exp(\tK)),
\label{Fdresc} \ee
where $G_d$ is the characteristic function when $\tK \mapsto 0$ in ${\cal B}[\td](u)$ [while all the other parameters in ${\cal B}[\td](u)$ are kept unchanged].
This allows us to rewrite the resummation expression (\ref{resFd}) in a simpler form (we rename $t'=t \exp(\tK)$ as new $t$)
\be
d(Q^2)_{\rm res} = \int_0^\infty \frac{dt}{t} G_d(t) {a}(t e^{-\tK} Q^2)
\label{dres2a}
\ee
and, as mentioned above, $G_d$ is the characteristic function when $\tK \mapsto 0$ in Eq.~(\ref{Btd5P})
\be
G_d(t) = \frac{1}{2 \pi i} \int_{u_0- i \infty}^{u_0 + i \infty} du {\rm B}[\td](u)|_{\tK \mapsto 0} \; t^u,
\label{GdinvMell}
\ee
where $-1 < u_0 <+1$.
\endgroup
Explicit evaluation gives:
\be
G_d(t) =   \Theta(1-t) \pi \left[ \frac{{\td}^{\rm IR}_1 t}{\Gamma(1-{k_1}) \ln^{{k_1}}(1/t)} + {\td}^{\rm IR}_2 t^2 \right]
+ \Theta(t-1) \pi \left[ \frac{{\td}^{\rm UV}_1}{t} + \frac{{\td}^{\rm UV}_2}{t^2} \right].
\label{Gd} \ee
It can be verified numerically that the resummation (\ref{dres2a}), using the expression (\ref{Gd}),  reproduces the correct perturbation series (\ref{dlpt}) when we Taylor-expand the coupling  $a(t e^{\tK} Q^2)$ ($ \equiv f[\ln(t e^{\tK} Q^2)]$) around any renormalisation scale $\ln \mu^2$, \textcolor{black}{i.e., the relations (\ref{tdnSR}) can be verified numerically.}

In practice, the pQCD coupling ${a}(Q^{'2})$ has (unphysical) Landau singularities at low positive $Q^{'2}$, thus in the integration  (\ref{dres2a}) we have to avoid them. We do this by the PV-type of regularisation
\be
d(Q^2)_{\rm res} = {\rm Re} \left[
\int_0^\infty \frac{dt}{t} G_d(t) {a}(t e^{-\tK} Q^2+ i \varepsilon) \right].
\label{dres2b}
\ee  
On the other hand, if we used an IR-safe coupling ${a}(Q^2) \mapsto {\A}(Q^2)$
that has no Landau singularities but practically coincides with ${a}(Q^2)$ at large $|Q^2| > \Lambda^2_{\rm QCD}$, such as the 3$\delta$AQCD coupling \cite{3dAQCD}, no additional regularisation would be needed
\be
d(Q^2)_{\rm res} = \int_0^\infty \frac{dt}{t} G_d(t) {\A}(t e^{-\tK} Q^2) .
\label{dres2c}
\ee
In the following, we will fit the OPE expression (\ref{BSROPE}), with terms up to dimension $D=4$ ($i=3$), to experimental data for BSR ${\overline \Gamma}_1^{p-n}(Q^2)$, where we evaluate the QCD canonical part $d(Q^2)$ with the renormalon-motivated resummation Eq.~(\ref{dres2b}). 

\section{Renormalisation scheme variation}
\label{sec:RSch}

First we notice from Table \ref{tabrenmod1} that in the $\MSbar$ scheme we have the two IR renormalon residues $\td^{\rm IR}_1$ and $\td^{\rm IR}_2$ with large values and opposite signs. This indicates that the two corresponding contributions to the canonical part of BSR, $d(Q^2)$, are large and have opposite signs, possibly even strong cancellations, which would be an unexpected behaviour. We can check this by performing the integration Eq.~(\ref{dres2b}), with $G_d$ of Eq.~(\ref{Gd}), term-by-term, cf.~Table \ref{tabrenmod2} (last row).

The expectation, based on arguments of \cite{Pin}, is that the leading IR renormalon contribution (${\rm IR}_1$: $\propto {\td}^{\rm IR}_1$) gives us the dominant contribution to $d(Q^2)$, and that the subleading IR contribution (${\rm IR}_2$: $\propto {\td}^{\rm IR}_2$) as well as the UV renomalon contributions (${\rm UV}_j$: $\propto {\td}^{\rm UV}_j$; $j=1,2$) all give numerically subdominant contributions to $d(Q^2)$. This is evidently not the case in our obtained renormalon-model resummation (\ref{dres2b}) in the $\MSbar$ scheme. Therefore, we will vary the renormalisation scheme (via the leading-scheme parameters $c_k \equiv \beta_k/\beta_0$; $k=2,3$) in such a way as to achieve the mentioned expected hierarchy of the four different renormalon contributions.

One may argue that, as the quantity $d(Q^2)$ must be renormalisation scale and scheme independent, so must be also the resummed results (\ref{dres2b}). The evaluated resummed quantity is exactly renormalisation scale independent, as mentioned above. However, it is not scheme independent (i.e., $\beta_j$-independent, where $j \geq 2$). This is so because the expression (\ref{Btd5P}) for ${\cal B}[\td](u)$ in general does not contain all the terms. Namely, the anomalous dimensiones corresponding to the three renormalons $u=2,-1,-2$ were taken to be zero (as are in the large-$\beta_0$ limit) and consequently the corresponding singularity structures there were taken to be simple poles. In reality, these terms are expected to be different from the simple poles. Further, for each such term $1/(p \mp u)^{\kappa}$ in ${\cal B}[\td](u)$ we expect to have subleading terms $\sim 1/(p \mp u)^{{\kappa}-1}$, and those terms were not included either, i.e., they were ``truncated out'' because of lack of information. The five-parameter ansatz (\ref{Btd5P}) is thus a somewhat simplified and truncated version, in which we were able to determine the parameters on the basis of the information about the pQCD perturbation series (\ref{dlpt}) truncated at $\sim {\ta}_5 \sim a^5$. For these reasons, we cannot expect that our resummed results (\ref{dres2b}) are invariant under the scheme variation. Therefore, we will have uncertainties of the extracted parameter values from scheme variation.

We then proceed in the following way. We vary the scheme, by varying the $c_2$ and $c_3$ scheme parameters (where $c_j \equiv \beta_j/\beta_0$). There are also other, more subleading, scheme parameters $\beta_j$ (or $c_j \equiv \beta_j/\beta_0$) ($j \geq 3$). For convenience, we will vary only the first two $c_j$ ($j=2,3$) and construct the beta-function with such $c_j$ which allows an explicit solution for the running coupling $a(Q^2; c_2, c_3)$ in terms of the Lambert function \cite{GCIK}. The corresponding beta-function $\beta(a)$ has a Pad\'e $[4/4](a)$ ('P44') form, \textcolor{black}{i.e., $\beta(a)$ is a coefficient of two polynomials of $a$, each of them of degree 4} 
\be
\frac{d a(Q^2)}{d \ln Q^2}  =
\beta(a(Q^2))
\equiv
- \beta_0 a(Q^2)^2 \frac{ \left[ 1 + a_0 c_1 a(Q^2) + a_1 c_1^2 a(Q^2)^2 \right]}{\left[ 1 - a_1 c_1^2 a(Q^2)^2 \right] \left[ 1 + (a_0-1) c_1 a(Q^2) + a_1 c_1^2 a(Q^2)^2 \right]} \ ,
\label{beta} \ee
where $c_j \equiv \beta_j/\beta_0$ and
\be
a_0  =  1 + \sqrt{c_3/c_1^3}, \quad 
a_1 = c_2/c_1^2 +   \sqrt{c_3/c_1^3} .
\label{a0a1} \ee
Expansion of this $\beta$-function up to $\sim a(Q^2)^5$ gives the expression (\ref{RGE2}) with $c_2$ and $c_3$. In Ref.~\cite{GCIK} it was shown that the RGE (\ref{beta}) has explicit solution in terms of the Lambert functions $W_{\mp 1}(z)$
\be
a(Q^2) = \frac{2}{c_1}
\left[ - \sqrt{\omega_2} - 1 - W_{\mp 1}(z) + 
\sqrt{(\sqrt{\omega_2} + 1 + W_{\mp 1}(z))^2 
- 4(\omega_1 + \sqrt{\omega_2})} \right]^{-1},
\label{aP44} \ee
where $\omega_1= c_2/c_1^2$, $\omega_2=c_3/c_1^3$,  $Q^2 = |Q^2| \exp(i \phi)$. The Lambert function $W_{-1}$ is used when $0 \leq \phi < \pi$, and $W_{+1}$ when $-\pi \leq \phi < 0$. The argument $z = z(Q^2)$ appearing in $W_{\pm 1}(z)$ is
\be
z \equiv z(Q^2) =
-\frac{1}{c_1 e}
\left( \frac{\Lambda_L^2}{Q^2} \right)^{\beta_0/c_1} \ .
\label{zexpr} \ee
Here, the scale $\Lambda_L$ we call the Lambert scale ($\Lambda_L \sim \Lambda_{\rm QCD}$). This scale is related with the strength of the coupling. We will call this class of schemes 'P44'. We recall that this coupling (\ref{aP44}), used in the resummation (\ref{dres2b}), has $N_f=3$ for all $Q^2$ in (\ref{dres2b}), because this resummation then corresponds to the perturbation expansions (\ref{dlpt}) or equivalently (\ref{dpt}) of BSR at low $Q^2$ where $N_f=3$. The coupling (\ref{aP44}) is determined by the value of $\alpha_s(M_Z^2; \MSbar)$ (which is at $N_f=5$) via the 5-loop $\MSbar$ RGE running and the corresponding 4-loop quark threshold relations at $Q^2 = \kappa {\bar m}_q$ (we took $\kappa=2$; ${\bar m}_b=4.2$ GeV and ${\bar m}_c=1.27$ GeV), and by changing the scheme from (5-loop) $\MSbar$ to the abovementioned scheme, at $Q^2 = (2 {\bar m}_c)^2 - 0$ (with $N_f=3$) by the known transformation (e.g., cf.~Eq.~(13) of Ref.~\cite{3dAQCD}).

For example, when we choose the P44 scheme with the $\MSbar$ values of $c_2$ and $c_3$, the parameters of ${\cal B}[\td](u)$ are those in the second row of Table \ref{tabrenmod1}, very close to the 5-loop $\MSbar$ case (first row there). The decomposition of the resummed $d(Q^2)$ expression (\ref{dres2b}), at $Q^2=3 \ {\rm GeV}^2$, in the separate renormalon contributions is given in the last row of Table \ref{tabrenmod2} in this (P44) $\MSbar$ case, which shows that the two IR contributions are large and with strong cancellation (over 90 \%). For this reason, we will consider such scheme as unacceptable in our approach.

We will confine ourselves to the schemes (P44-class) which give us the following type of contributions to $d(Q^2)$:
\begin{enumerate}
\item
IR $(u=1)$ contribution $d(Q^2)_{\rm IR1}$ is the dominant contribution.
\item
 The rescaling parameter $\tK$ in ${\cal B}[\td](u)$ is $|\tK|<1$.
\item
 IR $(u=2)$ contribution $d(Q^2)_{\rm IR2}$ will be restricted to:  $0< d(Q^2)_{\rm IR2} < d(Q^2)_{\rm IR1}$
\item
UV $(u=-2)$ contribution $d(Q^2)_{\rm UV2}$ should not be too large: $|{\td}^{\rm UV}_2| < 0.5$.  
\end{enumerate}
The conditions 2. and 3. turn out to be usually related: namely, if $ d(Q^2)_{\rm IR2} < 0$, we usually have  $|\tK|>1$, and $d(Q^2)_{\rm IR1}$ and $d(Q^2)_{\rm IR2}$ are large and with opposite signs and give strong cancellation. Taking into account these conditions, we obtain as acceptable (P44)-schemes those with
\be
c_2 = 9^{+2}_{-1.4}, \quad c_3= 20 \pm 15.
\label{c2c3} \ee
We note that when $c_2$ goes below the value $(9-1.4)$, the value of $|\tK|$ becomes suddenly large and we obtain strong cancellations of $d(Q^2)_{\rm IR1}$ and $d(Q^2)_{\rm IR2}$.

\begin{figure}[htb] 
\centering\includegraphics[width=80mm]{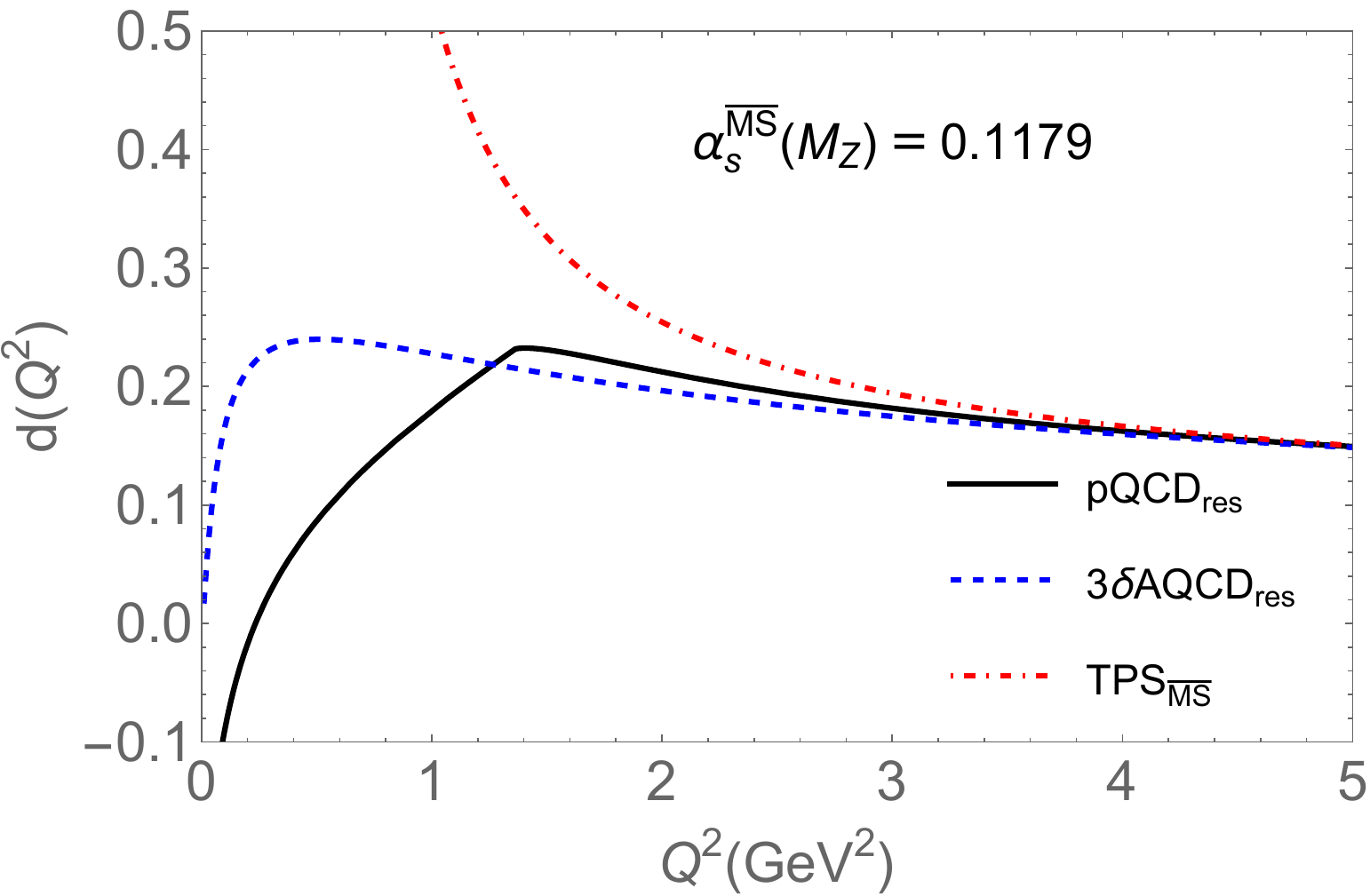}
\caption{\footnotesize The resummed canonical part of BSR, $d(Q^2)_{\rm res}$, according to Eq.~(\ref{dres2b}), in the 'P44' renormalisation scheme with $c_2=9.$ and $c_3=20.$, for $N_f=3$ ('${\rm pQCD}_{\rm res}$'). The strength of the coupling is determined by the choice $\alpha_s(M_Z^2;\MSbar)=0.1179$. The resummation with the corresponding 3$\delta$AQCD coupling is included, for comparison ('${\rm 3dAQCD}_{\rm res}$'). \textcolor{black}{Further, we include the evaluation as the truncated perturbation series (TPS) in powers of $a=a(Q^2)$ in $\MSbar$ scheme, up to $\sim a^5$ term and with the central value Eq.~(\ref{d4est}) of the corresponding coefficient, $d_4^{\MSbar}=1557.4$ ('${\rm TPS}_{\MSbar}$').}}
\label{FigdQ2}
\end{figure}

In Table \ref{tabrenmod2} we present the results for these P44-schemes, when $c_2$ and $c_3$ have the central values, or one of them varies to an edge value given in Eqs.~(\ref{c2c3}): the parameters of ${\cal B}[\td](u)$ and the decomposition of $d(Q^2)$ into the four contributions.
\begin{table}
\caption{\footnotesize The values of parameters of the five-parameter ansatz (\ref{Btd5P}) in the P44 renormalisation schemes, for various scheme parameters $c_2$ and $c_3$ covering the intervals Eq.~(\ref{c2c3}). Included are also the corresponding numerical values of the canonical BSR $d(Q^2)$ as well as its numerical decomposition to the four renormalon contributions, for the renormalon-motivated resummation (\ref{dres2b}), at $Q^2=3 \ {\rm GeV}^2$ (and $N_f=3$) and for $\alpha_s(M_Z^2;\MSbar)=0.1179$. The coefficient $d_4$ corresponds to the 5-loop $\MSbar$ coefficient value $d_4(\MSbar)= 1557.43$ as predicted by ECH.}
\label{tabrenmod2}
\begin{tabular}{rr|rrrrr|r|rrrr}
$c_2$ & $c_3$ & ${\tK}$ & ${\td_{1}^{\rm IR}}$ & ${\td_{2}^{\rm IR}}$ &  ${\td_{1}^{\rm UV}}$ & ${\td_{2}^{\rm UV}}$ & $d(Q^2)$ & $d(Q^2)_{\rm IR1}$ & $d(Q^2)_{\rm IR2}$ & $d(Q^2)_{\rm UV1}$ & $d(Q^2)_{\rm UV2}$ 
\\
\hline
\hline
9. & 20. & 0.450041 & 0.331813 & 0.231437  & -0.0809782 & -0.0964868 & 0.1816 & 0.1631 & 0.0597 & -0.0247 & -0.0164
\\
\hline
7.6 & 20. & 0.896252 & 0.210843 & 0.137235 & -0.158441 & 0.394581 &  0.1843 & 0.1196 & 0.0421 & -0.0551 & 0.0777
\\
11.0 & 20. & 0.12894 & 0.422324 & 0.301175 & -0.015641 & -0.477922 &  0.1807 & 0.1904 & 0.0701 & -0.0044 & -0.0752
\\
9. & 5. & 0.359327 & 0.474866 & -0.026115 & -0.080151 & -0.126694 &  0.1737 & 0.2243 & -0.0064 & -0.0236 & -0.0207
\\
9. & 35. & 0.484948 & 0.237256 & 0.431189 & -0.067963 & -0.133156 & 0.1888 & 0.1188 & 0.1143 & -0.0212 & -0.0232
\\
\hline
$c_2^{\MSbar}$ & $c_3^{\MSbar}$ & -1.83223 & 7.86652 & -14.9299 & -0.0444416 & -0.0776748 & 0.1632 & 1.9466 & -1.7675 & -0.0082 & -0.0076
\\
\hline
\end{tabular}
\end{table}
In Fig.~\ref{FigdQ2} we present the resummed values of the canonical BSR part $d(Q^2)$, Eq.~(\ref{dres2b}), for the considered central case of renormalisation scheme (P44 with $c_2=9.$ and $c_3=20.$) and with the strength of the coupling corresponding to the value $\alpha_s(M_Z^2;\MSbar)=0.1179$ (giving $\Lambda_{L}=0.2175$ GeV).
We can see in this Figure that the curve loses its expected monotonically decreasing behaviour for $Q^2 < 1.44 \ {\rm GeV}^2$. This occurs because for such low $Q^2$ the effects of the Landau singularities of the pQCD running coupling $a(t e^{-\tK} Q^2 + i \epsilon)$ in the integral (\ref{dres2b}) become significant.\footnote{In the considered case, the pQCD coupling $a(Q^{' 2})$ has Landau cut for $0 \leq Q^{'2} \leq 0.869 \ {\rm GeV}^2$.} Stated otherwise, the used renormalon-motivated resummation in the considered scheme starts failing at $Q^2 < 1.44 \ {\rm GeV}^2$ due to (unphysical) Landau singularities of the pQCD running coupling. In Fig.~\ref{FigdQ2} we included, for comparison, the results of resummation Eq.~(\ref{dres2c}) when the coupling $a(Q^2) \mapsto \mathcal{A}(Q^2)$ is holomorphic (i.e., without Landau singularities). We used a specific 3$\delta$AQCD coupling in miniMOM scheme, for the case $\alpha_s(M_Z^2;\MSbar)=0.1179$ and with the spectral function ${\rho}(\sigma) \equiv {\rm Im}\mathcal{A}(- \sigma - i \epsilon)$ with the threshold value $\sigma_{\rm thr} = M_1^2 = 0.150^2 \ {\rm GeV}^2$, for details see \cite{3dAQCD}.

\section{Fitting to the experimental data and conclusions}
\label{sec:fit}

\begin{figure}[htb] 
\begin{minipage}[b]{.49\linewidth}
\includegraphics[width=80mm,height=50mm]{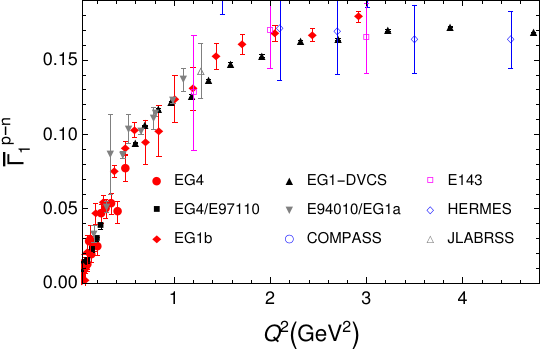}
\end{minipage}
\begin{minipage}[b]{.49\linewidth}
  \includegraphics[width=80mm,height=50mm]{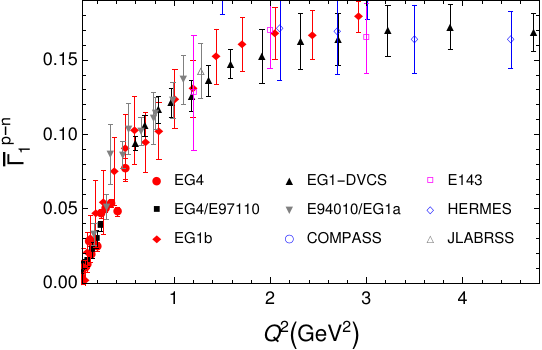}
\end{minipage} \vspace{12pt}
\caption{\footnotesize The measured results for the inelastic BSR ${\overline \Gamma}_1^{p-n}(Q^2)$ for different experiments, with the statistical (left Figure) and systematic (right Figure) uncertainties.}
\label{Figstasys}
\end{figure}

In Figs.~\ref{Figstasys} we present the numerical results for the inelastic BSR ${\overline \Gamma}_1^{p-n}(Q^2)$ from various experiments, with the statistical and systematic uncertainties .
We will perform the fit by using for $d(Q^2)$ the resummed expression (\ref{dres2b}) with the ($N_f=3$) pQCD coupling in the P44-renormalisation scheme with $c_2=9.$ and $c_3=20.$,  such that it corresponds to $\alpha_s(M_Z^2; \MSbar)=0.1179$ ($N_f=5$) which is the central value of the world average \cite{PDG2023}. The number of fit parameters will be either ${\bar f}_2$ or (${\bar f}_2$, $\mu_6$), i.e., we truncate the OPE (\ref{BSROPE}) at $D=2$ ($i=2$) or $D=4$ ($i=3$). We do not know which experimental uncertainties are correlated and which are not. The statistical uncertainties could be considered to be uncorrelated, but the correlations of the systematic uncertainties are expected to be considerable and difficult to estimate. Therefore, we follow the method of unbiased estimate \cite{Deuretal2022,PDG2020,Schmell1995}: a fraction of systematic uncertainty is added in quadrature to the statistical uncertainty, $\sigma^2(Q_j^2) = \sigma^2_{\rm stat}(Q_j^2) + k \sigma^2_{\rm sys}(Q_j^2)$, we consider then these $\sigma(Q_j^2)$ as uncorrelated, and we determine the fit parameters (${\bar f}_2$; or ${\bar f}_2$ and $\mu_6$) by minimising the corresponding $\chi^2/{\rm n.d.f.}$ for points in a chosen fixed interval $[Q^2_{\rm min}, Q^2_{\rm max}]$  (with $Q^2_{\rm max}=4.74 \ {\rm GeV}^2$). We continue adjusting the fraction parameter $k$ and minimising again, iteratively, until we obtain, when minimising, $\chi^2/{\rm n.d.f.}=1$. In practice, we always obtain $0 < k < 0.5$. The experimental uncorrelated uncertainty (exp.u.) of the extracted parameters is then obtained by the conventional methods (cf.~App. of Ref.~\cite{Bo2011}, App.~D of \cite{ACT2023}). The correlated experimental uncertainties (exp.c.) are then obtained by shifting the central experimental values ${\overline \Gamma}_1^{p-n}(Q^2_j)$ by $(1 - \sqrt{k}) \sigma_{\rm sys}(Q_j^2)$ up and down and reperforming the fit for these values.

We point out that the smaller the obtained value of $k$, the better the fit. It turns out that, in the above approach, the results depend considerably on the value of $Q^2_{\rm min}$ that we choose. We chose $Q^2_{\rm min}=1.71 \ {\rm GeV}^2$ for the fit with two parameters (${\bar f}_2$, $\mu_6$) for the following reasons. 1.) If we decrease $Q^2_{\rm min}$ to the adjacent lower neighbouring data points, the value of $k$ increases: from $k=0.162$  (for $Q^2_{\rm min}=1.71 \ {\rm GeV}^2$) to $k=0.172, 0.201$ (for $Q^2_{\rm min}=1.59,  1.50 \ {\rm GeV}^2$). If we decrease $Q^2_{\rm min}$ one step further, to $1.44 \ {\rm GeV}^2$, then we can see numerically that the evaluation of $d(Q^2)$ via Eq.~(\ref{dres2b}) is already on the border of applicability at such $Q^2$, due to the effects of the Landau singularities of $a(t e^{-\tK} Q^2 + i \epsilon)$ in the integral, cf.~Fig.~\ref{FigdQ2}. On the other hand, increasing $Q^2_{\rm min}$ above $1.71 \ {\rm GeV}^2$ to the upper neighbour $1.915 \ {\rm GeV}^2$, the value of $k$ increases (to $k=0.180$). If we increase $Q^2_{\rm min}$ even further, we obtain the results with strong cancellations between the $D=2$ and $D=4$ terms. For all these reasons, we choose $Q^2_{\rm min} = 1.71^{+0.205}_{-0.27} \ {\rm GeV}^2$, and the value of $k$ parameter \textcolor{black}{is $k=0.1621$.}

If the fit is performed only with one fit parameter (${\bar f}_2$), similar verifications give us $Q^2_{\rm min}=1.71^{+0.39}_{-0.27} \ {\rm GeV}^2$, and the value of $k$ parameter \textcolor{black}{is $k=0.1487$.}

  With the approach described above, we obtain the final result for the fits. For the two-parameter fit the result \textcolor{black}{is $k=0.1621$} and
\begingroup \color{black}
\bes
\label{resLTD24}
\bea
{\bar f}_2 & = & -0.160^{-0.007}_{+0.025}(c_2)^{+0.054}_{-0.039}(c_3)^{+0.044}_{-0.041}(\alpha_s)^{-0.012}_{+0.016}(d_4) \mp 0.043({\rm ren})
\nonumber\\ &&
^{+0.016}_{+0.119}(Q^2_{\rm min})\pm 0.160({\rm exp.u.}) \pm 0.297({\rm exp.c.}),
\label{bf2D24} \\
\mu_6 & = & +0.022^{+0.003}_{-0.008}(c_2)^{-0.013}_{+0.004}(c_3)^{-0.010}_{+0.008}(\alpha_s)^{+0.002}_{-0.003}(d_4) \mp 0.010({\rm ren})
\nonumber\\ &&
^{-0.006}_{-0.053}(Q^2_{\rm min})\pm 0.062({\rm exp.u.}) \mp 0.059({\rm exp.c.})
\; [GeV^4].
\label{mu6D24} \eea \ees
\endgroup
The quantity $\mu_6$ is in units of ${\rm GeV}^4$.
Here, the uncertainties at '$(c_2)$' and '$(c_3)$' come from renormalisation scheme variation Eq.~(\ref{c2c3}). The uncertainty at '$(\alpha_s)$' comes from the world average uncertainty $\alpha_s(M_Z^2; \MSbar)=0.1179 \pm 0.0009$ \cite{PDG2023}. The uncertainty at '$(d_4)$' comes from the variation of $d_4$ in such a way that the corresponding (5-loop) $\MSbar$ value $d_4^{\MSbar}$ varies according to Eq.~(\ref{d4est}). The uncertainty at '(ren)' is the renormalon uncertainty, it comes when in the evaluation of $d(Q^2)$, Eq.~(\ref{dres2b}), we add or subtract the same integral, but imaginary part (divided by $\pi$) instead of the real part  [$\pm (1/\pi) {\rm Im}(\ldots)$]. These are all the theoretical uncertainties.

The uncertainty at '($Q^2_{\rm min}$)' can be regarded as coming primarily from experimental uncertainties, and it originates from the variation $Q^2_{\rm min}=1.71^{+0.205}_{-0.27} \ {\rm GeV}^2$ as mentioned above. The experimental uncertainties at '(exp.u.)' and '(exp.c.)' were discussed in the previous paragraphs.

The one-parameter fit (${\bar f}_2$) gives, on the other hand, \textcolor{black}{$k=0.1487$ and}
\begingroup \color{black}
\bea
{\bar f}_2 & = & -0.107^{-0.001}_{+0.007}(c_2)^{+0.022}_{-0.029}(c_3) \pm 0.020(\alpha_s) \mp 0.009(d_4) \mp 0.067({\rm ren})
\nonumber\\ &&
^{+0.012}_{-0.029}(Q^2_{\rm min}) \pm 0.033({\rm exp.u.}) \pm 0.154({\rm exp.c.}),
\label{bf2D2} \eea
\endgroup
Here, the uncertainty  ($Q^2_{\rm min}$) comes from $Q^2_{\rm min}=1.71^{+0.39}_{-0.27} \ {\rm GeV}^2$ as mentioned above.

We note that we keep the (central) value of the parameter $k$ fixed under all the variations, except the variations of ($Q^2_{\rm min}$) where the amount of included experimental data points is varied and we require again $\chi^2/{\rm n.d.f}=1$.

\begingroup \color{black}
Furthermore, if we did not include the charm decoupling violation terms $\delta d(Q^2)_{m_c}$, Eq.~(\ref{deld}), in our analysis, then the results would change marginally: the central values in Eqs.~(\ref{resLTD24}) would change from $-0.160$ and $+0.022$ to $-0.165$ and $+0.023$, respectively, and in Eq.~(\ref{bf2D2}) the central value would change from $-0.107$ to $-0.108$, and all the uncertainties would remain practically unchanged. The $k$ parameter values would change from $0.1621$ and $0.1487$ to $0.1626$ and $0.1493$, respectively.
  \endgroup
  
The above results show that we have a competition between various theoretical uncertainties (which are in general moderate) and various experimental uncertainties of the extracted values. The latter uncertainties are large and are in general dominant over the theoretical uncertainties. The experimental uncertainties of the extracted parameter values have their origin, directly or indirectly, in the large statistical and systematic uncertainties of the BSR data points.

As mentioned earlier, the large experimental uncertainties of the data points make the deduction of the preferred value of $\alpha_s$ from the BSR data practically impossible, especially under the variation of $Q^2_{\rm min}$, and hence we used the world average data for $\alpha_s$.

\begin{figure}[htb] 
\begin{minipage}[b]{.49\linewidth}
\includegraphics[width=80mm,height=50mm]{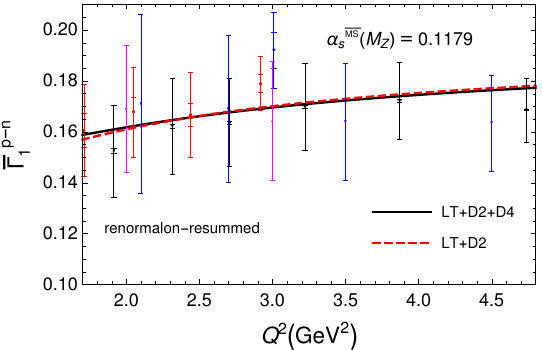}
\end{minipage}
\begin{minipage}[b]{.49\linewidth}
  \includegraphics[width=80mm,height=50mm]{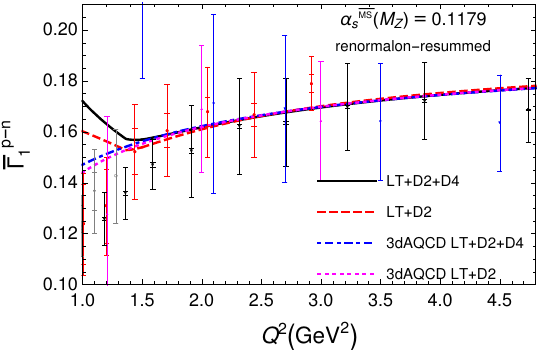}
\end{minipage} \vspace{12pt}
\caption{\footnotesize The theoretical OPE curves, when $D=2$ and $D=4$ terms are included, and when only $D=2$ term is included. The results are for the central values of ${\bar f}_2$ and $\mu_6$ obtained from the fit in the interval $Q^2_{\rm min} = 1.71 \ {\rm GeV}^2$, Eqs.~(\ref{resLTD24}) and (\ref{bf2D2}). The central renormalisation scheme is used (P44 with $c_2=9.$ and $c_3=20.$). The experimental data are included in the Figures. The left-hand Figure is for $Q^2 > 1.71 \ {\rm GeV}^2$ (the interval of fitting), and the right-hand Figure has extrapolation to the interval $Q^2 > 1.0 \ {\rm GeV}^2$. The right-hand Figure contains also the corresponding 3$\delta$AQCD curves, cf.~the text for explanations.}
\label{FigsBSRal01179}
\end{figure}

In Figs.~\ref{FigsBSRal01179} we present the obtained central fit theoretical curves (when truncation is made at $D=4$ and at $D=2$), i.e., when ${\bar f}_2$ and $\mu_6$ have the central values of Eqs.~(\ref{resLTD24}) and (\ref{bf2D2}), respectively. 
For comparison, we included in Figs.~\ref{FigsBSRal01179} the 3dAQCD curves (as in Fig.~\ref{FigdQ2}), where the fit for (${\bar f}_2, \mu_6$) or (${\bar f}_2$) was performed, as in pQCD case, for the interval with $Q^2_{\rm min}=1.71 \ {\rm GeV}^2$, and the same values of the $k$ parameter were used as in the pQCD case.\footnote{\textcolor{black}{In the used scheme [i.e., with $c_2=9.$ and $c_3=20.$, $N_f=3$, and $\alpha_s(M_Z^2,\MSbar;N_f=5)=0.1179$], the perturbative coupling $a(Q^2)$ has Landau singularities for $Q^2 < 0.9 \ {\rm GeV}^2$. Therefore, we used in the pQCD case in $\delta d(Q^2)_{m_c}$ and in the $D=2$ term for $a(Q^2)$ at $Q^2 < 1.5 \ {\rm GeV}^2$ simply the constant value $a(1.5) \approx 0.142$.}}

We can also apply, to the same interval of the data points, and for the same values of the $k$ parameter, the two-parameter and one-parameter fit when the canonical BRS part $d(Q^2)$ is evaluated as a simple truncated perturbation series (TPS), in $\MSbar$ scheme and for $\alpha_s(M_Z^2; \MSbar)=0.1179$. We have here additional theoretical uncertainties: the renormalisation scale dependence, and the truncation index ($N_{\rm tr}$) dependence.\footnote{This means that we truncate the TPS at the power at $a(\mu^2)^{N_{\rm tr}}$.}
We choose for simplicity for the renormalisation scale only the value $\mu^2=Q^2$.
Then we obtain for the two-parameter fit (${\bar f}_2$, $\mu_6$) with TPS a strong $N_{\rm tr}$ dependence; at $N_{\rm tr}=8$ we obtain small $\chi^2/{\rm n.d.f.} =0.891$, but there ${\bar f}_2 \approx -0.44$ and $\mu_6 =0.43 \ {\rm GeV}^4$ are both large and give significant cancellation effects between $D=2$ and $D=4$ BSR terms in the range $2 \ {\rm GeV}^2 < Q^2 < 3 \ {\rm GeV}^2$. For $N_{\rm tr} \geq 10$ we get $\chi^2/{\rm n.d.f.} > 4.$, i.e., very large. 
For the one-parameter fit (${\bar f}_2$) we obtain for all $N_{\rm tr} \geq 3$ the values $\chi^2/{\rm n.d.f.} >1$, and these values increase when $N_{\rm tr}$ increases.

For these reasons, we consider that our renormalon-motivated approach with the resummation Eq.~(\ref{dres2b}) is more reliable than the simpler TPS approach. As a consequence, Eqs.~(\ref{resLTD24}) and (\ref{bf2D2}), as well as Figs.~\ref{FigsBSRal01179}, represent the central results of our work. Furthermore, the presented work is an example of practical use of known renormalon information for an efficient evaluation (resummation) of the perturbation series of a spacelike observable in pQCD. 

\textcolor{black}{In this work, we did not consider models of BSR at very low $Q^2 \ll 1 \ {\rm GeV}^2$, such as expansions \cite{JeffL2} motivated on chiral perturbation theory or light-front holographic QCD (LFH) model \cite{LFH,LFHBSR}. We refer, for example, to \cite{ACKS} where such models of low-$Q^2$ BSR were included in the analyses there. Our analysis here was concentrated on (p)QCD approaches with the use of the QCD running coupling and OPE, and such methods fail are very low values of $Q^2$.}

The mathematica programs that were constructed and used in the calculations of this work, with the experimental data included, are available on the web page \cite{www}.

\begin{acknowledgments}
This work was supported in part by FONDECYT (Chile) Grants No.~1200189 and No.~1220095.
\end{acknowledgments}

\end{document}